\documentclass[10pt]{article} 
\usepackage{cite}
\usepackage{latexsym}
\usepackage{amsmath}
\usepackage{pstricks}
\usepackage{float}
\usepackage{pstricks}

\long\def\ca#1\cb{} 

 \def\outl#1{\par{\medskip\noindent\hspace*{.5cm}\bf
      \mathversion{bold}#1\mathversion{normal}\smallskip} }
 \def\xa{} \def\xb{}  

 \def\outl#1{}  \def\xa{} \def\xb{}  

\ca
 \def\outl#1{\par{\medskip\noindent\hspace*{.5cm}\bf
      \mathversion{bold}#1\mathversion{normal}\smallskip} }
 \long\def\xa#1\xb{}
 
\cb

\newcommand{\ad}{^\dagger }

\newcommand{\avg}[1]{\langle #1\rangle }
\newcommand{\becs}{\begin{cases}}
\newcommand{\bem}{\begin{matrix}}
 
\newcommand{\Blp}{\Bigl(}

\newcommand{\bra}[1]{\lgl#1|}
\newcommand{\Brp}{\Bigr)}

\newcommand{\dya}[1]{|#1\rgl\lgl#1|}

\newcommand{\encs}{\end{cases}}
\newcommand{\enm}{\end{matrix}}

\newcommand{\hf}{{\textstyle\frac{1}{2} }}

\newcommand{\hquad}{\mspace{8 mu}} 
\newcommand{\inp}[1]{\lgl#1|#1\rgl }

\newcommand{\ket}[1]{|#1\rgl }
\newcommand{\lbrk}[1]{\left\{\vrule height #1cm depth #1cm width 0pt\right.}

\newcommand{\lgl}{\langle } 
\newcommand{\lra}{\leftrightarrow }

\newcommand{\mte}[2]{\lgl#1|#2|#1\rgl }
\newcommand{\mted}[3]{\lgl#1|#2|#3\rgl }

\newcommand{\od}{\odot }
\newcommand{\ot}{\otimes }

\newcommand{\rgl}{\rangle }

\newcommand{\sgn}{\mathop{\rm sgn}}

\newcommand{\st}{\sqrt{2}}

\newcommand{\tm}{\times }
\newcommand{\Tr}{{\rm Tr}}

\newcommand{\vb}{\,|\,}

\newcommand{\AM}{{\mathcal A}}
\newcommand{\BM}{{\mathcal B}}
\newcommand{\CM}{{\mathcal C}}

\newcommand{\HM}{{\mathcal H}}

\newcommand{\JM}{{\mathcal J}}

\newcommand{\UM}{{\mathcal U}}
\newcommand{\VM}{{\mathcal V}}

\newcommand{\al}{\alpha }
\newcommand{\bt}{\beta }

\newcommand{\dl}{\delta }

\newcommand{\ep}{\epsilon}

\newcommand{\lm}{\lambda }
\newcommand{\Lm}{\Lambda }

\newcommand{\sg}{\sigma }

\begin{document}

\title{Quantum Locality}

\author{Robert B. Griffiths
\thanks{Electronic mail: rgrif@cmu.edu}\\ 
Department of Physics,
Carnegie-Mellon University,\\
Pittsburgh, PA 15213, USA}

\date{Version of 15 October 2010}

\maketitle

\xa
\begin{abstract}
 
  It is argued that while quantum mechanics contains nonlocal or entangled
  states, the instantaneous or nonlocal influences sometimes thought to be
  present due to violations of Bell inequalities in fact arise from mistaken
  attempts to apply classical concepts and introduce probabilities in a manner
  inconsistent with the Hilbert space structure of standard quantum mechanics.
  Instead, Einstein locality is a valid quantum principle:
  objective properties of individual quantum systems do not change when
  something is done to another noninteracting system.  There is no reason to
  suspect any conflict between quantum theory and special relativity.

\end{abstract}

\xb

	\section{Introduction}
\label{sct1}
\xa

\xb
\outl{Widespread idea that QM has nonlocal influences}
\xa

The opinion is widespread that quantum mechanics is nonlocal in the sense that
it implies the existence of long-range influences which act instantaneously
over long distances, in apparent contradiction to special relativity.  There
is a large and ever growing literature on this topic, so it is not practical
to give more than a few representative citations. See
\cite{Bll81,HyRd83,Bll90b,Albr92,PpRh94,Pty95,Glds96,SrSc97,Mdln02,Shmn04,%
ElDl05,dEsp06,Nrsn06,Shmn07,Ldsa08}
for some of the material advocating or defending nonlocality, and
\cite{Brkv07} for a helpful survey. There are certainly a substantial number
of dissenting voices, among them %
\cite{Fne82,Fne89,dMyn86,dMao94,dMyn02,Mrmn99,SmRv07,BrZk10}, 
but one is left with the impression that the proponents of nonlocality are
much more confident of their position than are their opponents---see, for
example, the recent exchange between Blaylock \cite{Blyl10} and Maudlin
\cite{Mdln10}, or a recent popular account claiming that quantum nonlocality
poses a serious problem for special relativity \cite{AlGl09}.  It is even
claimed, p.~18 of \cite{dEsp06}, that such influences are so well established
that not only are they a part of current quantum theory, but will necessarily
be part of any future theory that makes the same predictions about certain
experimental outcomes. While most of the discussion has taken place among
philosophers and physicists interested in the foundations of quantum
mechanics, one does not have to look very far in the physics research
literature or textbooks to find numerous claims that this or that experimental
confirmation of quantum predictions demonstrates the possibility of
``interaction free measurement,'' or the absence of ``local realism'' in the
quantum domain, or some similar idea tied to quantum nonlocality. See
\cite{DGao97,Kwao99,BnCS04s,WARZ05,DGrf05b,Ynao06,Grao07,MlLl08}
for a small but not atypical selection.

\xb
\outl{Nonlocal influences cannot transmit signals}
\xa

The widespread belief in the existence of such nonlocal effects seems a bit
surprising in view of theorems \cite{GhRW80,Brao00}, whose validity does not
seem to be in doubt, to the effect that these (supposed) quantum nonlocal
influences cannot be used to transmit signals or information. Thus they are
not detectable by any ordinary experimental test.  Normally (one would think)
this would immediately result in their being discarded into the dustbin of
metaphysics.  That this has not happened could reflect the unpalatable nature
of proposed alternatives that maintain locality, such as the idea that at the
atomic level physics no longer deals with or describes reality.  It seems
almost inevitable that if quantum mechanics is incompatible with ``local
realism'' (a term which in many minds is essentially synonymous with Bell
inequalities) but is still somehow local, then it is realism that must be
discarded, and one finds serious arguments to this effect \cite{CvFS07}.  But
if the world of atoms is unreal, what shall we say of the macroscopic objects
which most physicists think are composed of such atoms?  Are the things we see
around us (not to mention we ourselves) unreal?

\xb
\outl{Previous discussions not based on clear formulation of QM}
\xa

A serious problem with discussions of nonlocality, whether by advocates or
opponents, is that the arguments have not, up till now, been rooted in a
clear, unambiguous, and paradox-free formulation of quantum theory.  This
deficiency was pointed out in one of his last papers \cite{Bll90} by none
other than John Bell himself.  He was aware that such a formulation was not to
be found in the textbook treatments at the time he wrote, and the situation
has, unfortunately, not improved since then.  What students are taught in
courses in quantum mechanics are techniques for calculation, not the
fundamental principles of the subject.  The idea that a wave function
collapses when a measurement is carried out, to take a particularly pertinent
example, might represent something physical, in which case measurement
processes are good candidates for producing nonlocal influences, but it might
also be nothing but a technique for carrying out a calculation, something like
a change in gauge, with no reason at all to associate it with anything
physical, much less superluminal.  Conclusions based upon shaky or incomplete
or inconsistent formulations of quantum mechanics inherit, unfortunately,
these same uncertainties.

\xb
\outl{Present paper analyzes nonlocality from consistent QM approach}
\xa

The purpose of the present paper is to move beyond previous discussions by
employing a fully consistent quantum mechanical approach, aspects of which are
summarized below, to study the quantum (non)locality problem in a consistent
way that leads to quite definite conclusions.  We will argue, first, that
there are quite precise ways in which quantum mechanics can properly be said
to be nonlocal, in a manner which has no exact classical counterpart. Whether
in view of this one should call quantum theory itself ``nonlocal'' is a
different question. Second, we will argue that the supposed nonlocal
influences do \emph{not} exist, so the conflict with special relativity is
completely imaginary.  Third, we will establish on the basis of quantum
principles a strong statement of quantum \emph{locality}: the objective
properties of an isolated individual (quantum) system do not change when
something is done to another non-interacting system.

\xb
\outl{Hope is to raise the level of discussion}
\xa

As serious conceptual scientific issues are seldom resolved by a
straightforward analysis of arguments for or against a particular thesis
\cite{Khn70}, we are not so naive as to believe that the material presented
here will immediately result in unanimous consent with our conclusions.  We
do, however, hope to raise the level of discussion.  For too long the topic
has been debated in terms of ``measurement'' and ``free will'' and other
concepts which, as Bell put it \cite{Bll90}, should play no role in a
fundamental theory, and which at the very least make difficult the sort of
precise analysis appropriate to physics as a mature, exact science.  We claim
it is possible to do much better, and if the reader is capable of doing better
yet, so much the better.

\xb
\outl{QM applies universally}
\xa

Our starting point for addressing the issue of nonlocality is the belief that
quantum mechanics is at present our best fundamental theory of the mechanical
laws of nature and applies universally to systems of all sizes, microscopic or
macroscopic, from quarks to quasars, and under all conditions, whether it be
an atomic nucleus, a quantum optics experiment, or the center of the sun.  In
some circumstances classical mechanics is an excellent approximation to a more
fundamental quantum description, which is always available in principle, even
though it may be very awkward to construct it in detail.  In other
circumstances classical mechanics is inadequate and one must have resort to a
proper quantum description. Whenever classical and quantum mechanics are
significantly different it is the quantum description that best describes the
world as it is, and in this sense physical reality is quantum mechanical.  The
fact that the hydrogen atom cannot be described by the laws of classical
physics does not make it any less real; what this tells us is that real
reality is not classical reality.  The attitude expressed in \cite{ChCr93},
 that we should be willing to revise our familiar concepts of
reality to fit the teachings of quantum mechanics, has much to recommend it,
and is the approach taken in this paper.

\xb
\outl{``Local'' must be defined in Qm terms}
\xa

Although quantum and classical mechanics use many of the same words, such as
``energy'' and ``momentum'' and ``position'', the concepts are not exactly the
same.  Thus when discussing ``location'' or ``locality'' of things in the real
(quantum) world, one needs to employ well-defined quantum concepts that make
sense in terms of the quantum Hilbert space, and not simply import ideas from
classical physics with vague references to some ``uncertainty principle.''
Special relativity would have made little progress had Einstein restricted
himself to using ``time'' in precisely the same way it is employed in
pre-relativistic physics, and in the quantum domain one must be willing to
allow quantum mechanics itself to suggest a suitable concept of locality,
along with whatever limitations are needed to make it agree with the
mathematical structure of the theory.  Much of the confusion that attends
discussions of quantum (non)locality arises from an attempt to use classical
concepts in a quantum context where they do not fit, leading to internal
inconsistencies.

\xb
\outl{Paper assumes QM fundamentally stochastic, uses histories}
\xa

The present discussion is also based on the assumption that quantum mechanics
is \emph{fundamentally stochastic} or \emph{probabilistic}, and this
stochasticity is the way the world is, not something limited to the special
times or places at which measurements occur.  Consequently, quantum laws,
expecially those referring to dynamical processes, must be formulated in
consistent probabilistic terms.  At present the only known way to do this
within the framework of Hilbert space quantum mechanics (without additional
classical variables), with probabilities calculated from Schr\"odinger's
unitary time development (without additional artificial ``collapses'' added to
the dynamics) is to use the (consistent or decoherent) histories formulation
\cite{Grff84,Omns88,GMHr90,GMHr93,Grff96,Grff98,Omns99,Grff02c,GMHr07}, 
which will be employed throughout this paper.  The histories approach is
\emph{not} based upon the concept of measurement, and for this reason does not
suffer from the infamous ``measurement problem'' of quantum foundations
\cite{Wgnr63,BsSh96,Mttl98}.  Instead, it allows one to analyze measurements
using exactly the same formulation of quantum mechanics that applies to all
other physical processes. This analysis shows that when discussed in an
appropriate way, measurement outcomes for properly constructed pieces of
apparatus reveal properties possessed by measured systems \emph{before} the
measurement took place,%
\footnote{Measurements are discussed in considerable detail in Chs.~17 and 18
  of \cite{Grff02c}; the reader may also find helpful various toy models of
  measurements in earlier chapters.  For a more condensed discussion see
  \cite{Grff10}.} %
contrary to various claims found in textbooks.
Most of what follows can be understood without referring to the most
technically demanding part of the histories approach, the consistency
conditions, though these are used in the latter part of Sec.~\ref{sct5} and in
the proof in Sec.~\ref{sct6}.  In addition to the detailed discussion of
histories quantum mechanics in \cite{Grff02c}, the reader unfamiliar with the
this approach will find shorter introductions in
\cite{Grff09b,Hhnb10,Grff10}.

\xb
\outl{Outline of rest of paper}
\xa

\xb
\outl{1. Genuine Qm nonlocality}
\xa

Our argument aims to clear up the confusion surrounding quantum (non)locality
and put future discussions of this subject on a firm footing.  It involves
several steps.  The first is to discuss the sense in which quantum mechanics
allows nonlocal or entangled \emph{states}, and the precise way in which these
can be said to be nonlocal.  This genuine nonlocality, the topic of
Sec.~\ref{sct2}, needs to be clearly understood and clearly distinguished from
the notion of a nonlocal \emph{influence} of the sort mentioned above.  The
two have sometimes been confused, but they are in fact very different.

\xb
\outl{2. EPR and its proper interpretation }
\xa

Following this comes a discussion in Sec.~\ref{sct3} of the Bohm version
\cite{Bhm51s} of the famous Einstein, Podolsky and Rosen (EPR) situation
\cite{EnPR35}, starting with a macroscopic version involving colored slips of
paper, to which no one would ascribe any nonlocal influences, and going on to
argue that very much the same analysis applies in the case of measurements on
an entangled spin singlet pair.  To be sure, spin measurements allow for
additional possibilities that are not practical for slips of paper, and the
issue is how to interpret them and what conclusions can be drawn.  One can
come to paradoxical conclusions by ignoring the principles of quantum
mechanics, but the issue is fundamentally a local one; it is not nonlocal.

\xb
\outl{3. Qml derivation of Bell's inequality using a Cl approximation}
\xa

The majority of discussions of quantum nonlocality involve Bell inequalities 
\cite{CHSH69,Bll81,Bll90b} in some way or another.
All are agreed that quantum mechanics predicts large violations of the
Bell-CHSH inequality, and by now most physicists concede that experimental
tests support the quantum predictions. There remains a significant difference
of opinion as to what conclusions one should draw from this.  On one side are
those who claim that because locality is one of the assumptions that goes into
the derivation of the inequality, its violation proves that the quantum world
is nonlocal. Not just in the genuine sense discussed in Sec.~\ref{sct2}, but
that there exist nonlocal influences.  On the other side are those (including
the present author) who believe that other assumptions contradict quantum
principles, leading to an inequality that is incompatible with the quantum
nature of the world.
In the present paper we consider the Bell-CHSH inequality from two slightly
different perspectives.
First, in Sec.~\ref{sct4} we derive it by a method
which, while not exact from a quantum perspective can nevertheless be
justified as a good enough approximation for all practical purposes in the
case of golf balls.  One can then see how the derivation breaks down because
of noncommutation of operators in the domain where quantum effects can no
longer be ignored. The noncommutation has to do with \emph{local} operators.

\xb
\outl{4. Bell inequalities using FACT and INDE conditions}
\xa

Second, in Sec.~\ref{sct5} we take up the factorization and independence
conditions on probabilities involving a hidden variable (or variables),
denoted by $\lm$ in most treatments of the subject, which are often used in
deriving the inequality, and ask how they apply to a very simple microscopic
system represented by a quantum circuit involving four elements (all qubits in
the simplest case).  We insist that the probabilities be well defined and the
hidden variables be represented by quantum mechanical projectors, so as to
make the discussion relevant to the real (quantum) world.  Using the circuit
model one can ask about the possible existence of \emph{quantum} hidden
variables, properly defined quantum properties associated with subspaces of
the Hilbert space, which can play the role of $\lm$.  It turns out that
sometimes they exist and sometimes they do not exist, and this has very much
to do with local incompatibility issues, i.e., (non)commuting operators, and
nothing at all to do with mysterious nonlocal influences.  Our conclusion is
that this route to deriving Bell inequalities is not based on \emph{local
  realism}, as is often claimed, but instead on an assumption of
\emph{classical realism}.  This renders the Bell-CHSH inequality invalid for
drawing conclusions about the real (quantum) world, in particular its locality
or lack thereof.

\xb
\outl{5. Proof of Einstein locality}
\xa

Following these arguments we go on in Sec.~\ref{sct6} to establish the
positive result that objective properties of isolated individual systems do
not change when something is done to some other non-interacting system.  Thus
not only does the claim for nonlocal influences in the quantum domain rest
upon shaky arguments lacking a fundamental justification, it is also in flat
contradiction to a result based on consistent quantum arguments. Claims, such
as in \cite{Mdln02,AlGl09}, that supposed nonlocal influences make quantum
mechanics a threat to special relativity (or the reverse) are, consequently,
without foundation.
Various implications of our results are discussed in the concluding 
Sec.~\ref{sct7}.

\xb
\outl{Hardy's paradox, Stapp counterfactual not part of
  this paper}
\xa

In an article of modest length it is impossible to deal with all published
arguments claiming that quantum theory is beset with nonlocal influences and
in conflict with special relativity.  In particular we do not discuss those
based upon the GHZ \cite{GrHZ89,GHSZ90} or Hardy \cite{Hrdy92} paradoxes, nor
Stapp's counterfactual arguments.%
\footnote{See \cite{Stpp06} and references to earlier work found there.  Stapp
  agrees (private communication) that his arguments typically involve
  counterfactual elements.} %
Both Hardy's paradox and the problems associated with importing counterfactual
reasoning into the quantum domain are treated in some detail in Chs.~25 and
19, respectively, of \cite{Grff02c}, from a point of view similar to that in
this paper, and the conclusion regarding nonlocal influences is the same:
there is no evidence for them.  Also see \cite{Grff02b} with reference to
special relativity and quantum probabilities.

\xb
\section{Genuine Nonlocality}
\label{sct2}
\xa

\xb
\outl{Genuine quantum nonlocality = nonlocal states}
\xa

\xb
\outl{Nonlocal wavepacket}
\xa

Many errors contain a grain of truth, and this is true of the
mysterious nonlocal quantum influences. Quantum mechanics does
deal with states which are nonlocal in a way that lacks any precise classical
counterpart.  Understanding the nature of this genuine nonlocality will help
in identifying the misunderstandings which have given rise to the notion of
spurious  influences, so it is worth beginning with this topic. 
One does not need an entangled state, for the lack of this kind of locality is
already present in a wavepacket $\psi(x)$ for a single particle in one
dimension.  Imagine $\psi(x)$ spread out over an interval $x_1\leq x\leq x_2$;
a continuous function that is nonzero at any point inside this interval and
zero for all $x$ outside this interval.  The physical interpretation of
$\psi(x)$ depends on more than $|\psi(x)|^2$, for two wavepackets $\psi(x)$
and $\phi(x)$ may be such that $|\psi(x)|^2=|\phi(x)|^2$, and yet be
orthogonal to each other, which means that they represent distinct quantum
properties.  Thus the physical meaning of $\psi(x)$ depends on relative phases
or phase differences at different points in space.  The same is true in
relativistic quantum mechanics where the phase differences are at spacelike
separate points.  Similarly, entangled states of two spin-half ions in a trap,
or two photons traveling away from a crystal where they were produced by down
conversion, can properly be said to be nonlocal.

\xb
\outl{Localization within an interval. Incompatibility of $\psi(x)$ and
  locality}
\xa

But in what sense are they nonlocal? Let us discuss this for $\psi(x)$. Since
it vanishes outside the interval $[x_1,x_2]$ it is \emph{localized} to (or on)
this interval, but inside the interval itself it is \emph{delocalized} in the
following precise sense.  Let $[\psi] =\dya{\psi}$ be the projector onto the
subspace of the Hilbert space of square integrable functions that contains
$\psi(x)$, and $X(x',x'')$ the projector onto the interval $[x',x'']$,
\begin{equation}
  X(x',x'')\phi(x) = 
\begin{cases}
\phi(x) &\text{ for $x'\leq x \leq x''$}\\ 
0 &\text{ elsewhere.}
\end{cases}
\label{eqn1}
\end{equation}
Since by assumption $\psi(x)$ is nonzero throughout the interval from $x_1$ to
$x_2$, the projectors $[\psi]$ and $X(x',x'')$ will fail to commute when $x'$
or $x''$ is in the interior of $[x_1,x_2]$, i.e., if $x_1<x'<x_2$ or if $x_1 <
x'' < x_2$.  That is to say, the property%
\footnote{As used in this article the term ``property'' as applied to a
  quantum system always corresponds to some (closed) subspace of the Hilbert
  space or, equivalently, the projector onto this subspace, and is to be
  distinguished from a ``physical variable'' represented by a Hermitian
  (self-adjoint) operator.  Thus the eigenspace of the Hamiltonian $H$
  corresponding to a particular eigenvalue $\ep$ represents the property
  $H=\ep$.  See the discussion in Chs.~4 and 5 of \cite{Grff02c}.} %
of being ``local,'' located within the interval $[x',x'']$, is incompatible in
the quantum sense of noncommuting projectors with the property of being ``in
the state $\psi$'' if the interval $[x',x'']$ is chosen
injudiciously. Informally, the particle cannot be localized more precisely
than the support of its wave function.  Because the issue is one of
noncommutation of operators, or incompatibility of the corresponding physical
properties, we are dealing with a feature of the quantum world
which lacks any precise analog in, and which cannot be reduced to, classical
physics.  The essential point is \emph{not} that a measurement of position
will not yield a predictable result.  While this is true, it is also true in
the case of a classical particle described in classical statistical mechanics
with a probability distribution.  What is new in the quantum case is
noncommutation. This needs to be recognized as such, not dismissed with a
vague reference to an uncertainty principle.

\xb
\outl{Such nonlocality typical of entangled states including spin singlet}
\xa

This sort of nonlocality is typical of entangled states when the entanglement
involves objects or situations which are spatially separated from each other,
as in the case of the spin singlet state
\begin{equation}
  \ket{\psi_0} = \Blp\ket{+}_\AM\ket{-}_\BM - \ket{-}_\AM\ket{+}_\BM\Brp/\st,
\label{eqn2}
\end{equation}
of two spin-half, spatially separated particles, where we assume that
$\ket{+}$ corresponds to $S_x=+1/2$ in units of $\hbar$, and $\ket{-}$ to
$S_x=-1/2$, and the subscript indicates which particle.  (This is the
well-known example Bohm \cite{Bhm51s} used to elucidate the EPR problem.)  One
cannot ascribe a property to the spin of particle $\AM$, such as $S_{\AM
  x}=+1/2$, if the combined system has the property of being in the state
$\ket{\psi_0}$.  The relevant projectors do not commute, so the local property
is \emph{incompatible} with the property represented by the entangled state.
In this quite precise \emph{quantum mechanical} sense the state \eqref{eqn2}
can be said to be \emph{nonlocal} when the two particles are in different
locations as specified by the spatial parts of the total wave function.

\xb
\outl{Noncommuting projectors common in QM; nonlocality one manifestation}
\xa

Nonlocality in this sense is just one manifestation of the quite general
situation that arises when properties are defined using projectors on the
quantum Hilbert space instead of subsets of points in the classical phase
space, and different properties may correspond to projectors that do not
commute, the hallmark of quantum in contrast to classical physics.  Properties
which are incompatible do not have a conjunction that makes physical sense.
If $P_1$ and $P_2$ are two commuting projectors, then $P_1 P_2 = P_2 P_1$ is a
projector representing the property ``$P_1$ AND $P_2$.''  But if $P_1 P_2 \neq
P_2 P_1$, neither product is a projector, and there is no natural way
(consistent with ordinary logic) of defining a corresponding ``$P_1$ AND
$P_2$'' property.  See the discussion in Ch.~4 of \cite{Grff02c}.

\xb
\outl{QM is also nonenergetic and nonmomentous if it is nonlocal }
\xa

Quantum incompatibility manifests itself in various ways which are not directly
connected with local properties.  For example, it is incorrect to ascribe a
property to the spin of the electron in a hydrogen atom in its ground state,
of which \eqref{eqn2} is the spin part, even though the electron and proton
are not spatially separated but lie on top of each other (to the extent
possible in the quantum world).  Nontrivial unitary time dependence is
described by states which are ``nonenergetic'' in the sense that the projector
onto such a state at some given time fails to commute with the Hamiltonian,
which is to say with the projectors making up its spectral
decomposition. Similarly, wavepackets of finite extent are ``nonmomentous'' as
well as being ``nonlocal'' in the sense described above.

\xb
\outl{Contrast genuine Qm nonlocality with nonlocal influences}
\xa

Genuine quantum nonlocality thus refers to descriptions of quantum systems in
terms of their local properties, or perhaps the incompatibility of such
properties with the specified quantum state.  By contrast, the notion of
\emph{nonlocal influences}, also called spooky action at a distance,
superluminal influences, etc., refers to a (supposed) \emph{dynamical} effect:
what is done to a certain quantum system here can have an instantaneous
influence on a second system located far away, be it at the other end of the
laboratory or (in principle) the other side of the galaxy.  Nonlocal
properties, which are very much part of the quantum world, do not by
themselves imply the presence or absence of nonlocal influences.  However, as
we shall see, various \emph{claims} that quantum theory contains nonlocal
influences can reflect a failure to take adequate account of certain
\emph{local} manifestations of quantum incompatibility.

\xb
\section{Correlations Not Causes}
\label{sct3}
\xa

\xb
\outl{Colored slips of paper: Charlie, Alice, Bob}
\xa

Charlie in Chicago takes two slips of paper, one red and one green, places
them in two opaque envelopes, and after shuffling them so that he himself does
not know which is which, addresses one to Alice in Atlanta and the other to
Bob in Boston, both of whom know the protocol Charlie is following.  Upon
receipt of the envelope addressed to her Alice opens it and sees a red slip of
paper.  From this she can immediately conclude that the slip in Bob's envelope
is green, whether or not Bob has already opened his envelope, or will ever do
so. Her conclusion is not based on a belief that opening her envelope to
``measure'' the color of the slip of paper has some magical long-range
influence on the color of Bob's slip.  Instead it employs statistical
reasoning in the following way.

\xb
\outl{Alice's reasoning upon seeing red slip}
\xa

Before Alice opens the envelope she (or Bob or Charlie) can assign
probabilities to the various situations as follows:
\begin{align}
  \Pr(A=R,B=R) & = \Pr(A=G,B=G) = 0,
\notag\\ 
 \Pr(A=R,B=G) &= \Pr(A=G,B=R) = 1/2 ,
\label{eqn3}
\end{align}
where $A=R$ means a red slip in Alice's envelope, $B=G$ a green slip in Bob's,
etc., and $\Pr()$ refers to the joint probability distribution. From
the usual rule for conditional probabilities, $\Pr(C\vb D) = \Pr(C,D)/\Pr(D)$ 
it follows that
\begin{equation}
  \Pr(B=R\vb A=R) = 0,\quad \Pr(B=G\vb A=R) = 1,
\label{eqn4}
\end{equation}
and this is the conditional probability distribution that Alice uses to
infer the color of Bob's slip knowing that the one in her envelope is red.
One could say that she uses the outcome of her observation to ``collapse'' the
initial probability distribution \eqref{eqn3} onto the conditional probability
distribution \eqref{eqn4}.  The colors of the two slips of paper, the
ontological situation in the physical world, is not at all affected by Alice's
``measurement.''  It is her \emph{knowledge} of the world that changes, in a
way which we do not find at all surprising.  The ``collapse,'' if that is what
one wishes to call it, refers to a method of reasoning, not a physical effect.

\xb
\outl{Spin singlet state; Alice and Bob measurements}
\xa

Next consider a situation in which Charlie at the center of the laboratory
pushes a button, and one member of a pair of spin-half particles initially in
the entangled state \eqref{eqn2} is sent towards Alice's apparatus at one end
of the building, while the other is simultaneously sent towards Bob's
apparatus at the other end.  If Alice measures the $x$ component of spin of
her particle and the apparatus indicates $A=1$ corresponding to $S_x=1/2$
before the measurement took place---let us assume that Alice is a competent
experimentalist who has designed and built a piece of apparatus which can do a
measurement of this sort%
\footnote{In talks given by experimental particle physicists one hears all
  sorts of references to trajectories and spins of elementary particles before
  they are measured.  That this talk is perfectly justified from the
  perspective of quantum mechanics has been known for a long time; see Sec.~2
  of \cite{Grff84}, and for more details Chs.~17 and 18 of
  \cite{Grff02c}. Competent experimentalists ignore the nonsense they were
  taught in their introductory quantum mechanics courses about measurements
  producing results out of nowhere.  The quantum foundations community should
  do the same.}%
---what can she say about $S_x$ for Bob's
particle?  The chain of probabilistic inference is identical to that discussed
earlier for colored slips of paper, though, as we shall see, certain details
must be discussed with greater care.

\xb
\outl{Quantum sample spaces for spin half.  Incompatible possibilities}
\xa

In order to use probabilities in quantum theory in a consistent way it is
necessary to specify, as in every application of probability theory, a
\emph{sample space}%
\footnote{See, for example, p.~4 of \cite{Fllr68} or p.~6 of \cite{DGSc02}.} %
of mutually exclusive possibilities: one and only one of which occurs in any
particular experimental run.  Two incompatible properties of a quantum system
cannot both be included in the same sample space since they are not mutually
exclusive.  This makes choosing a sample space for spin half particles more
subtle than the corresponding problem for slips of paper, even though in
principle the two problems (viewed from a quantum perspective) are similar.
The other novelty for spin half is that there are various different
incompatible sample spaces one may wish to consider, whereas for slips of
paper there is in practice no need to consider alternatives to the usual
quasiclassical framework:
\cite{GMHr90,GMHr93,GMHr07} and Sec.~26.6 of \cite{Grff02c}.  

\xb
\outl{Choice of sample space does not influence reality}
\xa

Note, incidentally, that choosing the sample space is part of setting up a
stochastic or probabilistic theoretical model of the situation, and the
reality being modeled is no more influenced by the physicist's choice of a
sample space than would the flow of traffic on the Pennsylvania Turnpike be
influenced by someone in the Department of Transportation setting up a
stochastic model thereof.  Using one sample space rather than another will
determine what properties the physicist is able to discuss, but it does not
influence these properties. A helpful but limited analogy---no classical
picture can accurately portray all aspects of the quantum world---is that of a
photographer choosing to photograph a mountain from the north rather than from
the south.  Different perspectives yield different possibilities for obtaining
information.%
\footnote{The reader may find it helpful to look at various examples in Ch.~18
  of \cite{Grff02c} in order to see how different sample spaces can be used
  in the context of quantum measurements.} %

 \xb 
\outl{Decompositions of $I$ as Qm sample spaces.  The $\{[\pm]\ot[\pm]\}$
  choice} 
\xa

Quantum sample spaces are always (projective) decompositions of the identity
$I$ on the quantum Hilbert space,
and for the problem under discussion it is useful to use the following 
four projectors
\begin{equation}
  [+]_\AM\ot [+]_\BM,\hquad [+]_\AM\ot [-]_\BM,\hquad [-]_\AM\ot [+]_\BM,
\hquad  [-]_\AM\ot [-]_\BM, 
\label{eqn5}
\end{equation}
where $[+] =\dya{+}$, $[-] =\dya{-}$, and the subscripts are particle
labels. The projector $[+]_\AM\ot[-]_\BM$ corresponds to $S_{\AM x}=+1/2$, and
$S_{\BM x}=-1/2$. The four projectors are mutually orthogonal, so represent
mutually exclusive events, and add up to the identity $I_\AM\ot I_\BM$ on the
spin space of the two particles, thus covering all possibilities. We are
interested in the corresponding probabilities at a time $t_1$ when the
particles are close to but before they reach the measuring apparatuses, the
counterpart of a time before the envelopes are opened in the case of slips of
paper.

\xb
\outl{Joint probabilities for $S_{\AM x}$ and $S_{\BM x}$}
\xa

The probabilities associated with the events in \eqref{eqn5} can be calculated
in the following way.  Assume Charlie has constructed an apparatus which will
reliably produce the spin state $\ket{\psi_0}$ in \eqref{eqn2} at some time
$t_0 < t_1$, and that the quantum unitary time development operator
$T(t_1,t_0)$ from time $t_0$ to $t_1$ for the spin degrees of freedom is the
identity, so
\begin{equation}
\ket{\psi_1} = T(t_1,t_0)\ket{\psi_0} = \ket{\psi_0}.  
\label{eqn6}
\end{equation}
One may then use the Born rule
\begin{equation}
  \Pr(Q) = \mte{\psi_1}{Q}
\label{eqn7}
\end{equation}
to assign a probability to any of the properties $Q$ in \eqref{eqn5} at the
time $t_1$, with the result 
\begin{align}
  \Pr(S_{\AM x}&=+\hf,\;S_{\BM x}=+\hf)  = 
 \Pr(S_{\AM x}=-\hf,\;S_{\BM x}=-\hf) = 0,
\notag\\ 
 \Pr(S_{\AM x}&=+\hf,\;S_{\BM x}=-\hf) =
 \Pr(S_{\AM x}=-\hf,\;S_{\BM x}=+\hf)  = 1/2.
\label{eqn8}
\end{align}

\xb
\outl{Reasoning using these probabilities}
\xa

The probabilities in \eqref{eqn8} are the analogs of those in \eqref{eqn3},
and by using them Alice can apply exactly the same form of statistical
reasoning as in the case of slips of paper.  If her observation (apparatus)
indicates that the particle arriving at her end of the laboratory had $S_x =
+1/2$ just before the measurement took place, as reflected in the measurement
outcome, she can infer that the particle arriving at Bob's end of the
laboratory had $S_x =-1/2$ just before his measurement took place. Thus the
mode of statistical reasoning is \emph{identical} for spins and slips of paper
once an appropriate sample space has been adopted and the probabilities
calculated by the rules of quantum dynamics for the system under
consideration.
At this point a number of comments are in order.

\xb
\outl{Comment 1. Spins discussed before measurements, because measurements may
  disturb them}
\xa

1. Why discuss spins of particles at a time \emph{before} any
measurements took place?  Because measurements of this type can have violent
effects on the properties measured.  In a Stern-Gerlach apparatus, for
example, there is no reason to suppose that the spin of the particle after the
measurement, when the particle may have been destroyed or stuck to a glass
plate, is the same as before.  In the case of measurements which preserve
a property, so that it is the same after the measurement as before, one can of
course also consider the spin state after the measurement, but in general this
is not possible.

\xb
\outl{Comment 2. Measurement outcomes reflect measured properties}
\xa

2. Why suppose that the macroscopic outcome of Alice's reflected a property
which the particle had before the measurement took place?  It is possible in
principle to analyze the action of the apparatus in fully quantum mechanical
terms, and show that for an appropriately constructed apparatus, having
properties which should be satisfied if it is constructed by a competent
experimentalist, the $S_x$ values before the measurement and the pointer
position after the measurement are appropriately correlated: the latter
indicates the former.  See Chs.~17 and 18 of \cite{Grff02c}. Of course this is
not the sort of argument which will convince every sceptic, but it is the sort
of reasoning which physicists apply to experimental results all the time.  The
issue is analogous to what one meets in astronomy: why does the image on the
CCD accurately reflect the light coming in from the sky?  One needs to
understand how the apparatus works.

\xb
\outl{Comment 3.  Alice could have measured $S_z$ instead of $S_x$}
\xa

3. Could not Alice have measured $S_z$ instead of $S_x$, by using a modified
apparatus?  Yes, and in that case the measurement outcome using the
appropriately constructed apparatus would have provided no information about
the value of $S_x$, and she would be able to say nothing at all about $S_x$
for Bob's particle.  But couldn't she have used the outcome to say something
about $S_z$ for Bob's particle?  Yes, but to do so would have meant employing
a sample space different from, and incompatible with, that in \eqref{eqn5}.
Alice has a choice in determining what it is she measures, and this choice
determines what she can say on the basis of the measurement outcome about
properties of Bob's particle.  Later, Sec.~\ref{sct5b}, we will construct a
toy model of what it means for an experimentalist to make a choice, and
analyzing this model can aid understanding.  But in no case does Alice's
choice of which measurement to do have any influence whatsoever on any
property of Bob's particle: this will be demonstrated in Sec.~\ref{sct6}.

\xb
\outl{Comment 4. Wave function as a pre-probability}
\xa

4. An important conceptual point relates to the role of $\ket{\psi_1}$ as
defined by unitary time development in \eqref{eqn6} and used in \eqref{eqn7}.
When employing the sample space \eqref{eqn5} at the time $t_1$, $\ket{\psi_1}$
cannot be considered a physical property of the particle system at this time,
because $[\psi_1]$ does not commute with the projectors in \eqref{eqn5}.
Instead it enters \eqref{eqn7} as a calculational tool, a device for computing
probabilities, a pre-probability in the terminology introduced in Sec.~9.4 of
\cite{Grff02c}.  Failing to distinguish a ket as a calculational device from a
ket as a physical property (to be precise, the property corresponds to the ray
containing this ket, or the projector onto this ray) is a source of
considerable conceptual confusion.

\xb
\outl{Comment 5. Calculating probabilities using wave function collapse}
\xa

5. Cannot the same probabilities be calculated by collapsing a wave function?
They can, and wave function collapse should be considered a
\emph{calculational device}, a means of computing probabilities, often
conditional probabilities, that can be obtained equally well by other methods
with less spectacular names.  The careless use of ``collapse'' ideas has led
to all sorts of conceptual difficulties and paradoxes which can be avoided if
one is careful about setting up sample spaces and calculating quantum
probabilities according to well-defined fundamental principles.  On the other
hand, collapse can be a very useful computational tool.  Suppose that Alice
has measured $S_x=+1/2$ for her particle.  She can then use $\ket{\psi_1}$ to
compute a ``conditional'' ket $\ket{-}_\BM =\st\bra{+}_\AM\ket{\psi_1}$, in a
somewhat awkward use of Dirac notation, for the spin state of Bob's particle,
thought of as a pre-probability.  This conditional ket provides a fast way to
calculate probabilities of outcomes of Bob's measurements if they are made in
some basis other than the $[\pm]_\BM$ of \eqref{eqn5}.  An equally good
approach, though it takes a trifle longer, is to carry out the ``collapse'' by
taking the partial trace over $\AM$ of the operator $([+]_\AM \ot I_\BM)
[\psi_1]$. The point is that ``collapse'' is one way of carrying out a
calculation involving a stochastic model, following the standard rules of
probability theory, and has nothing to do with any mysterious influence from
one end of the laboratory to the other.

\xb 
\outl{Comment 6. Alice assigning different state after measurement than
  before is no surprise} 
\xa

6. Both in the case of slips of paper and in the case of spin half, Alice can,
quite properly, assign a different probabilistic description to the world
after she knows the outcome of her observations than she can before.  This
difference could be manifested in assigning a different wave function (as in
the previous paragraph) or a different density operator, or in various other
ways.  Since quantum mechanics has long been regarded, at least in practice,
as a stochastic theory, this should not come as a surprise.  Unfortunately,
conditional probabilities are not properly treated in quantum textbooks.  Wave
function collapse is used for this purpose, but it is then confused with a
physical effect rather than being properly identified as a calculational tool.

\xb 
\outl{Comment 7. Alice could have measured a different component of spin}
\xa

7. Suppose Alice has her apparatus set up to measure $S_x$ and obtains
some value, which she, as a competent experimentalist, believes to be the
value possessed by this component of the spin angular momentum of the particle
before the measurement took place.  But she could instead have measured $S_z$,
and in that case she would have reached the conclusion that $S_z$ had a
particular value.  Thus not only did $S_x$ have the value actually measured,
but it must also have had some value of $S_z$, the value that \emph{would have
  been} measured in the counterfactual world in which the apparatus was set up
to measure $S_z$.  Is this not paradoxical?

\xb
\outl{This reasoning resembles EPR; has counterfactual character}
\xa

The reader familiar with the original EPR paper \cite{EnPR35} will immediately
recognize in this the sort of argument by which they concluded that quantum
mechanics is incomplete, except that now it is entirely a ``local'' matter
having to do with just one particle.  The weak point in such reasoning is its
counterfactual character; see the discussion of quantum counterfactuals in
Sec.~19.4 of \cite{Grff02c} for an analysis of what can go wrong. The problem
is always some violation of the single framework rule, some mixing of
arguments from incompatible sample spaces. Thus an argument, as in the
preceding paragraph, which starts off with $S_x$ and ends up discussing $S_z$
for the same particle at the same instant of time cannot be embedded in a
single sample space, and is thus an attempt to apply reasoning appropriate to
classical physics in a context in which it doesn't work.

\xb
\outl{Summary of section}
\xa

To summarize this section before going on to additional topics. The
correlations of the Bohm-EPR sort are similar to and can be understood using
probabilistic reasoning of the same sort involved in the analogous situation
involving colored slips of paper, where no one would ever suggest the
existence of nonlocal influences. The analysis in both cases has to be carried
out using a properly defined sample space.  For slips of paper viewed quantum
mechanically there is only a single quasiclassical sample space that needs to
be taken into account in practice. Although the quantum physicist can very
well imagine other possibilities (e.g., with slips of different colors in some
macroscopic or Schr\"odinger-cat superposition) they are of no practical
interest.  By contrast, alternative incompatible sample spaces are relevant to
possible laboratory experiments on spin half particles (or correlated
photons), so choosing the sample space---a choice made by the physicist, not
by some ``law of nature''---is not automatic, and care must be exercised not
to mix conclusions based on incompatible sample spaces.

\xb
\section{Golf Balls}
\label{sct4}
\xa

\xb
\outl{Definitions of $A_a$, $B_b$, $C(a,b)$, CHSH inequality for golf balls}
\xa

Consider two spinning golf balls $\AM$ and $\BM$, and let $L_{\AM a}$ and
$L_{\BM b}$
denote the $a$ and $b$ components of their spin angular momentum (e.g., 
$a$ might be $x$ or $z$). Further let
\begin{equation}
  A_a = \sgn(L_{\AM a}),\quad  B_b = \sgn(L_{\BM b})
\label{eqn9}
\end{equation}
be the algebraic signs, $\pm1$ or $0$ of these components.  In a probabilistic
setting in which $L_{\AM a}$ and $L_{\BM b}$ are random variables one can
define a correlation function
\begin{equation}
  C(a,b) = \avg{A_a B_b},
\label{eqn10}
\end{equation}
where $\avg{\cdot}$ denotes the average taken with respect to the joint
probability distribution.  The Bell-CHSH inequality \cite{CHSH69,Bll81,Bll90b}
\begin{equation}
  |C(a,b) + C(a,b') + C(a',b) - C(a',b')| \leq 2,
\label{eqn11}
\end{equation}
where $a$ and $a'$ are denote two (in general different) components of the
spin of particle $\AM$, and $b$ and $b'$ are similarly defined,
is a consequence of the following straightforward argument.

\xb
\outl{Random variable $W$ lies in $[-2,2]$}
\xa

Define the random variable
\begin{equation}
  W = A_a B_b +A_a B_{b'} +A_{a'} B_b -A_{a'} B_{b'}
\label{eqn12}
\end{equation}
From \eqref{eqn10} and the linearity of $\avg{\cdot}$ one sees that the left
side of \eqref{eqn11} is $|\avg{W}|$.  Since $W$ depends on each of the four
quantities $A_a$, $A_{a'}$, $B_b$, and $B_{b'}$ in a linear manner it follows
that if each of these is confined to the range $[-1,+1]$, the maximum and
minimum of $W$ must occur when each is $+1$ or $-1$, and then by looking at
the various possibilities one sees that $W$ itself is confined to the interval
\begin{equation}
  -2\leq W\leq 2.
\label{eqn13}
\end{equation}
Since the
average of a quantity that always lies between $-2$ and $+2$
falls in the same range, we have demonstrated the correctness of \eqref{eqn11}.
Thus \eqref{eqn11} follows at once from the definitions in
\eqref{eqn9} and \eqref{eqn10} through a straightforward application of
standard probability theory.  It has nothing to do with how the golf balls
were set spinning in the first place, or how the components were measured (if
they were measured), and nothing to do with locality or location.  To be sure,
it is hard to imagine two golf balls on top of each other at the same
location, but it is easy to imagine that $\AM$ and $\BM$ are the same golf
ball at a single location but at two different times, in which case
\eqref{eqn11} will again be valid.

\xb
\outl{Quantum derivation using $J_{\AM a}$ and $J_{\BM b}$}
\xa

This derivation was carried out using \emph{classical} physics. Let us see
what happens when we try and ``quantize'' it by replacing $L_{\AM a}$ and
$L_{\BM b}$ with the corresponding quantum operators $J_{\AM a}$ and $J_{\BM
  b}$ (in units of $\hbar$), with commutation relations
\begin{equation}
  [J_{\AM x},J_{\AM y}] = i J_{\AM z}, \quad
  [J_{\BM x},J_{\BM y}] = i J_{\BM z}, \quad
  [J_{\AM x},J_{\BM y}] = 0,
\label{eqn14}
\end{equation}
and so forth. 
Replace \eqref{eqn9} with
\begin{equation}
 \hat  A_a = P_{\AM +1} - P_{\AM -1},\quad
 \hat B_b = Q_{\BM +1} - Q_{\BM -1}.
\label{eqn15}
\end{equation}
where $P_{\AM +1}$ is a projector onto the subspace of the
Hilbert space spanned by eigenvectors of $J_{\AM_a}$ with positive eigenvalues, 
$P_{\AM -1}$ onto those with negative eigenvalues, and the $Q$'s are
similarly defined using $J_{\BM b}$. Let $\hat W$ be the operator obtained by
putting hats on the symbols on the right side of \eqref{eqn12}. It
is Hermitian, since the $\hat A$'s commute with the $\hat B$'s, and therefore
if its eigenvalues fall within the interval \eqref{eqn13} the quantum average
$\avg{W}$ will fall within the same range, and we again arrive at 
\eqref{eqn11}.  

\xb
\outl{Quantum derivation of CHSH fails because of noncommuting operators}
\xa

Alas, \eqref{eqn13} no longer holds if $W$ is replaced by $\hat W$.  What is
wrong with the derivation immediately preceding \eqref{eqn13} in the quantum
case, i.e., if we put hats on the $A$ and $B$ symbols?  This is seen most
easily by considering the extreme case where $\AM$ and $\BM$ are spin-half
particles.  Suppose that $a=z$ and $a'=x$, so that $\hat A_a$ and $\hat
A_{a'}$ become the noncommuting Pauli operators $\sg_z$ and $\sg_x$, and also
let $b=z$ and $b'=x$.  One sees that the spectra of the individual operators
lies in the range $[-1,1]$, the same as with the classical random variables
assumed previously.  However, because the operators no longer commute the
``subtraction'' trick by which the last term in \eqref{eqn12} in effect
``cancels'' one of the preceding terms no longer suffices to restrict the
spectrum of $\hat W$ to the range \eqref{eqn13}.  To summarize, while
classical reasoning can often be applied in the quantum domain and still yield
correct results, there are occasions in which it fails and one has to employ
the correct quantum tools.

\xb
\outl{Breakdown has nothing to do with nonlocality}
\xa

Note that the failure has to do with \emph{noncommuting operators}: were it
the case that $\hat A_a\hat A_{a'} =\hat A_{a'} \hat A_a$ the argument leading
up to \eqref{eqn13} would still be correct, though one would want to reword it
in the language of operators rather than classical random variables.  Thus we
have to do with an argument which is valid in the classical domain, but whose
quantum counterpart breaks down.  Note that this the breakdown has no
connection with anything nonlocal.  The $\hat A$ operators commute
with the $\hat B$ operators, but not among themselves.  To be sure, in order
to violate \eqref{eqn11} we have to consider a situation in which $\hat A_a$
fails to commute with $\hat A_{a'}$ \emph{and} $\hat B_b$ fails to commute
with $\hat B_{b'}$.  But this is no sign of nonlocality; it merely shows that
physicists sometimes get away with careless arguments.

\xb
\outl{But CHSH is an excellent \emph{classical} approximation for large $J$}
\xa

Despite the foregoing remarks one can still justify the application of
\eqref{eqn11} to golf balls through the following chain of reasoning. 
Let us take the commutation relation \eqref{eqn14}
and divide it by the total angular momentum
quantum number $J$ so that it becomes
\begin{equation}
  [L_{\AM x},L_{\AM y}] = i L_{\AM z}/J
\label{eqn16}
\end{equation}
in terms of ``normalized'' dimensionless operators $L_{\AM x}=J_{\AM x}/J$,
etc., which are then of order 1 in a typical situation.  As $J$ for a spinning
golf ball is of the order of $10^{30}$, we see that noncommutativity is in
practice not likely to be of much concern. Thus while the derivation of
\eqref{eqn11} is, strictly speaking, wrong even for golf balls, it is at least
plausible that there are circumstances in which we are justified in saying
that it holds as an excellent \emph{classical approximation}, valid for all
practical purposes when discussing macroscopic objects, or even spinning
molecules if they are not too small.  And as with all classical
approximations, we are not surprised if \eqref{eqn11} breaks down in a
situation where those approximations are no longer valid.  To be sure, showing
that classical physics emerges in an appropriate limit from quantum physics is
not altogether straightforward, and the approximate commutation exhibited in
\eqref{eqn16} when $J$ is large is only the beginning of the process.  To do
it properly requires the introduction of an appropriate coarse-graining using
quasi-classical frameworks---see the brief remarks in Sec.~26.6 of
\cite{Grff02c}, and see \cite{GMHr93} for more details.  Some care is needed,
for even when $J$ is very large it is possible to exhibit Bell-like
inequalities violated by quantum mechanics \cite{Mrmn80,GrMr83}.  That,
however, only adds further strength to the case being made here: inequalities
of this sort belong to the domain of classical, not quantum physics

\xb
\outl{Further problem due to sums of noncommuting operators}
\xa

There is actually another serious defect with the attempt to derive
\eqref{eqn11} in the quantum case along the lines suggested above, by placing
hats on the operators that appear in \eqref{eqn12}.  The difficulty is that
\eqref{eqn12} is then a sum of noncommuting Hermitian operators.  While it is
formally true that the sum of the averages is equal to the average of the
sums, this hides a conceptual difficulty: what is the average that one is
talking about?  Students of quantum theory learn that placing a Hermitian
operator inside angular brackets brings good marks on the examination, but
there is empirical evidence \cite{SnBC06} that they do not understand
what they are doing. (We may count it fortunate that no similar research has
been carried out on their instructors!)  A proper use of probability theory
requires that a sample space be specified, and while in quantum theory this is
often implicitly taken to be defined by the eigenspaces of the Hermitian
operator in question, that is obviously problematic when taking the sum of
noncommuting operators which may have no eigenspaces in common.  Sometimes an
effort is made to evade this problem by claiming that the only probabilities
being referred to are the outcomes of measurements. But then one must pay
serious attention to how measurement outcomes are related to measured
quantities, as in the discussion in the following section. 

\xb
\outl{Bell-CHSH inequality one of many results not always true in Qm world}
\xa

In summary, the Bell-CHSH inequality is one of many results in
\emph{classical} statistical physics which do not always hold in the real
(quantum) world.  It can be derived from quantum principles provided one makes
approximations which in many situations will be valid for all practical
purposes in the case of macroscopic objects.  Nothing in the classical or
quantum derivations gives the least suggestion of any nonlocal influences.

\xb
\section{Hidden Variables}
\label{sct5}
\xa

\xb
\subsection{Factorization and independence}
\label{sct5a}
\xa

\xb
\outl{Factorization (locality) and independence conditions}
\xa

Bell's inequality can be derived provided the probabilities for a certain set
of random variables satisfy two equations.  The first is often called
``locality'' but we will refer to as the \emph{factorization condition}:
\begin{equation}
  \Pr(A,B\vb a,b,\lm) = \Pr(A\vb a,\lm)\Pr(B\vb b,\lm).
\label{eqn17}
\end{equation}
The second is the \emph{independence} condition
\begin{equation}
 \Pr(\lm\vb a,b) = \Pr(\lm).  
\label{eqn18}
\end{equation}
In a formal sense we are dealing with five random variables $A, B, a, b, \lm$,
and the probabilities in these equations are marginal probabilities obtained
from a ``master'' distribution $\Pr(A,B,a,b,\lm)$. Summing over some
variables yields the marginal distribution, e.g. $\Pr(A,B,a) = 
\sum_{b,\lm} \Pr(A,B,a,b,\lm)$, for those which remain, and conditional
probabilities are defined as usual.

\xb
\outl{Physical meanings of $A$, $B$, $a$, $b$, $\lm$}
\xa

In the present application $A$ and $B$ are to be thought of as macroscopic
\emph{measurement outcomes}, taking values between $-1$ and $+1$, on two
separated but perhaps correlated systems $\AM$ and $\BM$, while $a$ and $b$
refer to the \emph{types} of measurements carried out; e.g., settings of knobs
that determine which properties of $\AM$ and $\BM$ are measured in order to
yield the outcomes $A$ and $B$.  The interpretation of the \emph{hidden
  variable} (or variables, but it is quite adequate to use a single discrete
variable) $\lm$ is much less clear, and that is the essential topic to be
discussed.  One usually thinks of it as some sort of microscopic state
associated with the systems to be measured before the measurement takes place.
Hopefully its significance will become clearer in the following discussion.
Its only role in the derivation of Bell's inequality \eqref{eqn11} is that it
\emph{exists} in such a way that \eqref{eqn17} and \eqref{eqn18} are
satisfied.

\xb
\outl{Combine FACT and INDE to get BELL equation}
\xa

\xb
\outl{Definition of $C(a,b)$; derivation of CHSH}
\xa

The first step in obtaining \eqref{eqn11} is to combine
\eqref{eqn17} and \eqref{eqn18} to yield
\begin{equation}
  \Pr(A,B\vb a,b) = \sum_\lm \Pr(A\vb a,\lm) \Pr(B\vb b,\lm) \Pr(\lm),
\label{eqn19}
\end{equation}
which we leave to the reader as an exercise in probability theory; note that
it is not a consequence of \eqref{eqn17} alone, but also requires
\eqref{eqn18}.
Next---compare with \eqref{eqn10}---define
\begin{equation}
C(a,b) = \sum_{A,B} AB\Pr(A,B\vb a,b)
  = \sum_{A,B} AB\Pr(A\vb a,\lm)\Pr(B\vb b,\lm) \Pr(\lm)
 = \sum_\lm  A_a(\lm) B_b(\lm) \Pr(\lm)
\label{eqn20}
\end{equation}
where
\begin{equation}
  A_a(\lm) := \sum_A A \Pr(A\vb a,\lm),\hquad
  B_b(\lm) := \sum_B B \Pr(B\vb b,\lm),
\label{eqn21}
\end{equation}
are quantities lying in the interval $[-1,1]$, given that both $A$ and $B$ are
in this range. Consequently, using the same
argument that leads to \eqref{eqn13}, we conclude that
\begin{equation}
  W(\lm) = A_a (\lm)B_b(\lm) +A_a (\lm)B_{b'}(\lm) 
    +A_{a'}(\lm) B_b (\lm)-A_{a'}(\lm) B_{b'}(\lm)
\label{eqn22}
\end{equation}
lies in the interval $[-2,2]$. Thus multiplying \eqref{eqn22} by $\Pr(\lm)$
and summing over $\lm$ leads to the Bell-CHSH inequality \eqref{eqn11}.

\xb
\outl{Interpretation, identification of $\lm$ using golf balls}
\xa

Since, as noted earlier, \eqref{eqn11} does not hold in general in the quantum
world, it follows that there are circumstances in which either \eqref{eqn17}
or \eqref{eqn18} or both fail, which is to say one can find no quantum hidden
variable $\lm$ such that both conditions are satisfied.  Before going
further it is worth remarking why one might expect both of these to hold in
circumstances in which classical physics gives an excellent approximation to
quantum mechanics.
Imagine that the golf balls of Sec.~\ref{sct4} have been shot out in opposite
directions from a machine that has first set them spinning in opposite
senses about some randomly chosen axis.  Next, measurements of different
components of angular momentum are carried out by competent experimentalists,
so the outcomes $A$ and $B$ reflect properties of the golf balls just before
the measurements took place.  One would expect these measurement outcomes to
show correlations: $\Pr(A,B\vb a,b)$ as determined experimentally would not be
equal $\Pr(A\vb a)\Pr(B\vb b)$.  But if $\lm$ is the hidden variable $(\vec
L_\AM,\vec L_\BM)$ corresponding to the \emph{actual} angular momenta of each
of the golf balls just \emph{before} measurements take place, one would expect
\eqref{eqn17} to hold---each of the terms is either 0 or 1---for a properly
constructed apparatus.  And \eqref{eqn18} is the plausible assumption that
$\vec L_\AM$ and $\vec L_\BM$ are functions of the initial preparation of the
golf balls and the air resistance they have encountered on their way to the
measurement apparatus, but are not influenced by the settings of apparatus
knobs, which could have been selected by a random number generator at the very
last moment.

\xb
\subsection{Quantum circuit}
\label{sct5b}
\xa

\xb
\outl{Circuit introduced}
\xa

\begin{figure}[h]
$$
\begin{pspicture}(-1.7,-5.8)(4.9,3.5) 
\newpsobject{showgrid}{psgrid}{subgriddiv=1,griddots=10,gridlabels=6pt}
\def\lwd{0.035} 
\def\lwb{0.05}  
\def\lwdsh{0.020} 
\def\rcp{0.25}  
\def\rdot{0.10} \def\rodot{0.12} 
\def\rdet{0.35}  
\psset{
labelsep=2.0,
arrowsize=0.150 1,linewidth=\lwd}
\def\circ{%
\pscircle[fillcolor=white,fillstyle=solid]{0.4}}
\def\cnt{\pscircle(0,0){\rcp}\psline(-\rcp,0)(\rcp,0)\psline(0,-\rcp)(0,\rcp)}
\def\cnotg(#1,#2,#3){%
\psline(#1,#2)(#1,#3)\rput(#1,#2){\cnt}\pscircle*(#1,#3){\rdot}}
\def\csqug(#1,#2,#3){%
\psline(#1,#2)(#1,#3)\rput(#1,#2){\squ}\pscircle*(#1,#3){\rdot}}
\def\detect{
\psarc[fillcolor=white,fillstyle=solid](0,0){\rdet}{-90}{90}
\psline(0,-\rdet)(0,\rdet)}
\def\ddetect{
\psarc[fillcolor=white,fillstyle=solid](0.2,0){\rdet}{-90}{90}
\rectg(0,-\rdet,0.2,\rdet)}
\def\dot{\pscircle*(0,0){\rdot}} 
\def\odot{\pscircle[fillcolor=white,fillstyle=solid](0,0){\rodot}} 
\def\dput(#1)#2#3{\rput(#1){#2}\rput(#1){#3}} 
\def\rectc(#1,#2){%
\psframe[fillcolor=white,fillstyle=solid](-#1,-#2)(#1,#2)}
\def\rectg(#1,#2,#3,#4){
\psframe[fillcolor=white,fillstyle=solid](#1,#2)(#3,#4)}
\def\squ{\rectc(0.35,0.35)}
\def\tmline(#1,#2,#3)#4{
\rput(#1,#2){\psline[linestyle=dashed,linewidth=\lwdsh](0.0,-0.3)(0.0,#3)}
\rput(0,-0.4){\rput[t](#1,#2){#4}}}
\def\vertdash{\psline[linestyle=dashed,linewidth=\lwdsh](0.0,-0.3)(0.0,1.5)}
                \def\figa{
\psline(0,0)(4,0)
\psline(0,1)(4,1)
\psline(0,2)(4,2)
\psline(0,3)(4,3)
\psline(3,0)(3,1) \rput(3,0){\dot}
\psline(3,2)(3,3) \rput(3,3){\dot}
\dput(3,1){\squ}{V}
\dput(3,2){\squ}{U}
\rput(4,0){\detect}
\rput(4,1){\detect}
\rput(4,2){\detect}
\rput(4,3){\detect}
\tmline(0.2,-.1,3.5){$t_0$}
\tmline(1,-.1,3.5){$t_1$}
\tmline(2.3,-.1,3.5){$t_2$}
\tmline(3.6,-.1,3.5){$t_3$}
\rput[b](0.6,0.1){$b$}
\rput[b](0.6,1.1){$\BM$}
\rput[b](0.6,2.1){$\AM$}
\rput[b](0.6,3.1){$a$}
\rput[l](4.5,0){$b$}
\rput[l](4.5,1){$B$}
\rput[l](4.5,2){$A$}
\rput[l](4.5,3){$a$}
\rput[r](-.1,0){$\ket{\phi_b}$}
\rput[r](-.1,3){$\ket{\phi_a}$}
\rput[r](0,1.5){$\lbrk{0.5}$}
\rput[r](-.4,1.5){$\ket{\Phi}$}
\rput(-1.5,1.5){(a)}
               }
                \def\figb{
\psline(0,0)(1.5,0)
\psline[linestyle=dashed,linewidth=\lwb](1.5,0)(4,0)
\psline[linestyle=dashed,linewidth=\lwb](1.5,3)(4,3)
\psline(0,1)(4,1)
\psline(0,2)(4,2)
\psline(0,3)(1.5,3)
\psline[linestyle=dashed,linewidth=\lwb](3,0)(3,1) \rput(3,0){\dot}
\psline[linestyle=dashed,linewidth=\lwb](3,2)(3,3) \rput(3,3){\dot}
\dput(3,1){\squ}{V}
\dput(3,2){\squ}{U}
\rput(1.5,0){\detect}
\rput(4,1){\detect}
\rput(4,2){\detect}
\rput(1.5,3){\detect}
\tmline(0.2,-.1,3.5){$t_0$}
\tmline(1,-.1,3.5){$t_1$}
\tmline(2.3,-.1,3.5){$t_2$}
\tmline(3.6,-.1,3.5){$t_3$}
\rput[b](0.6,0.1){$b$}
\rput[b](0.6,1.1){$\BM$}
\rput[b](0.6,2.1){$\AM$}
\rput[b](0.6,3.1){$a$}
\rput[l](4.5,0){$b$}
\rput[l](4.5,1){$B$}
\rput[l](4.5,2){$A$}
\rput[l](4.5,3){$a$}
\rput[r](-.1,0){$\ket{\phi_b}$}
\rput[r](-.1,3){$\ket{\phi_a}$}
\rput[r](0,1.5){$\lbrk{0.5}$}
\rput[r](-.4,1.5){$\ket{\Phi}$}
\rput(-1.5,1.5){(b)}
               }
\rput(0,0){\figa} \rput(0,-5){\figb}
\end{pspicture}
$$
\caption{%
Quantum circuit with (a) all measurements after $t_3$; (b) measurements on
ancillaries before $t_2$ determine the later measurement types.}
\label{fgr1}
\end{figure}
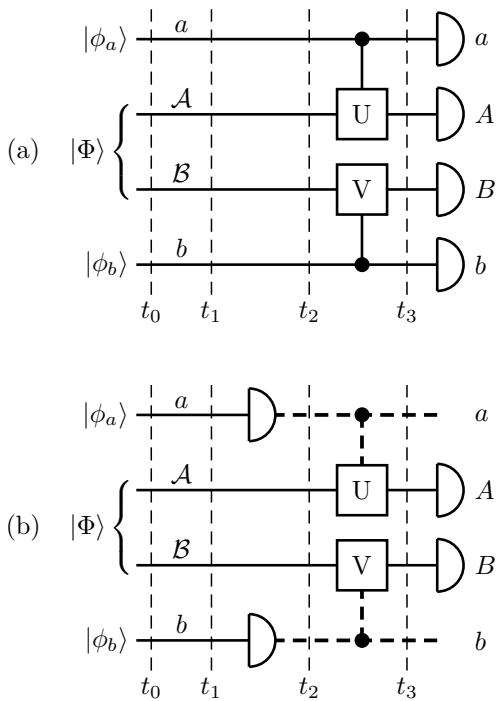

What goes wrong with \eqref{eqn17} or \eqref{eqn18} in the quantum world?
Rather than attempt an abstract discussion it is helpful to set up a specific
model, shown in Fig.~\ref{fgr1} as a quantum circuit in which time increases
from left to right.%
\footnote{Such circuits are widely used in quantum information theory; see,
  e.g., Ch.~4 of \cite{NlCh00} or Chs.~1 and 2 of \cite{Mrmn07},} %
The four horizontal lines represent $D$-state
quantum systems or qudits, but all the essential ideas are present for the
$D=2$ qubit case, think of spin-half particles.  They undergo interactions
shown by boxes connected to vertical lines.  The total
Hilbert space
\begin{equation}
  \HM = \HM_a \ot\HM_\AM \ot\HM_\BM \ot\HM_b 
\label{eqn23}
\end{equation}
is the tensor product of those for the individual particles. Different times
$t_j$ in the order $t_0 < t_1 < t_2 < t_3$ are indicated in the figure by
dashed vertical lines.

\xb
\outl{Ancillary systems, controlled gates, time development}
\xa

In Fig.~\ref{fgr1}(a) $a$ and $b$ are ancillary systems used to generate
different types of measurements, and kets in the corresponding orthonormal
(standard or computational) bases are denoted by $\ket{a}$ and $\ket{b}$, with
$a$ and $b$ taking on the values $1, 2,\ldots$. (Note that (quantum) computer
scientists prefer to start the count at 0 rather than 1, a matter of
indifference for the present discussion.) The square boxes $U$ and $V$
connected to vertical lines represent the gates, or unitary operations,
\begin{equation}
  \UM = \sum_a \dya{a}\ot U^{(a)},\quad 
 \VM = \sum_b \dya{b}\ot V^{(b)}
\label{eqn24}
\end{equation}
carried out on $\HM_a\ot\HM_\AM$ and $\HM_b\ot\HM_\BM$, respectively.  Hence
the total unitary time development from time $t_2$ to $t_3$ corresponds to the
operator
\begin{equation}
  T(t_3,t_2) = \UM\ot\VM,
\label{eqn25}
\end{equation}
while $T(t',t)$ on any other time interval is
determined by the usual composition rule
\begin{equation}
  T(t_l,t_j) = T(t_l,t_k)T(t_k,t_j)
\label{eqn26}
\end{equation}
and the fact that $T(t_1,t_0)$ and $T(t_2,t_1)$ are simply the identity $I$ on
$\HM$.

\xb
\outl{Measuring apparatuses and measured quantities}
\xa

The $D$ symbols in Fig.~\ref{fgr1} denote measurements and the dashed lines at
later times in (b) denote the ``classical'' measurement outcomes. In a fully
quantum description the measurement apparatus must itself be described in
quantum terms; see, for example, Chs.~17 and 18 of \cite{Grff02c}. Including
the apparatus Hilbert space(s) produces no problem in principle, but doing so
adds nothing (except technical complications) to the following discussion.  We
shall simply assume, as in most quantum information discussions, that the
measurements are ideal: there is a one-to-one correspondence between the
pointer positions, represented schematically by dashed lines, and the prior
values of the measured qubits (or qudits).  Thus $\ket{a=1}$ at $t_3$ in
Fig.~\ref{fgr1}(a) or $t_1$ in (b) leads to a measurement outcome or pointer
position which can also be denoted by $a=1$, and likewise for the $b$ qubit.
In the case of $A$ and $B$ we adopt a slightly different convention in which
the qubit kets are labeled by $\pm 1$, thus kets $\ket{A=1}$ and $\ket{A=-1}$
will lead to measurement outcomes $A=1$ and $A=-1$, since this is what we have
been using when discussing Bell's inequality.

\xb
\outl{Measurements on ancillaries before and after the controlled gates}
\xa

Figure~\ref{fgr1}(b) differs from (a) in that the measurements on the
ancillary systems $a$ and $b$ have been carried out at an earlier time, just
after $t_1$, and the classical or macroscopic outcomes are used to determine
which unitary operation to apply to $\AM$ or $\BM$.  It is well known%
\footnote{See \cite{GrNi96} and the ''Principle of deferred measurement'' on
  p.~186 of \cite{NlCh00}.} %
that the circuits (a) and (b) yield exactly the same measurement statistics,
the same joint probability distribution $\Pr(A,B,a,b)$. For the present
discussion the advantage of (b) is that it provides a quantum model for a
situation in which the \emph{type} of measurement, that is, the $a$ and $b$
value, is determined at the very last instant $t_2$ before the measurement
takes place through what is in effect flipping a quantum coin.  By choosing
the state $\ket{\phi_a}$ to be a superposition of the $\ket{a}$ states, the
value of $a$ at $t_2$, which determines which type of measurement will be
carried out on $\AM$, is decided at the very last instant; similarly for the
$\BM$ side.  As we shall see, this guarantees that \eqref{eqn18} is satisfied
for the hidden variables of interest to us, allowing us to focus on
\eqref{eqn17}.

\xb
\outl{Controls determine the type of measurement or measurement basis}
\xa

But in what sense does the value of $a$ determine the type of measurement?
Applying a unitary transformation $U^{(a)}$ to the particle $\AM$ and then
measuring it at time $t_3$ in the standard basis is completely equivalent to
measuring it at time $t_2$ in a different basis consisting of the states
obtained by applying ${U^{(a)}}\ad$ to the standard basis $\ket{A=\pm1}$. This
principle is by now well accepted in the quantum information community, and
the reader not familiar with it is invited to check it by working out some
examples.

\xb
\outl{Formula for $\Pr(A,B,a,b)$ }
\xa

Assuming an initial state at time $t_0$ of the form, 
\begin{equation}
  \ket{\Psi_0} = \ket{\Phi}\ot\ket{\phi_a}\ot\ket{\phi_b}
\label{eqn27}
\end{equation}
and using the circuit in Fig.~\ref{fgr1}(a), we can calculate the joint
probability distribution of interest at time $t_3$ by using the Born rule,
\begin{equation}
  \Pr(A,B,a,b) = \mte{\Psi_3}{\Blp [a]\ot[A]\ot[B]\ot[b] \Brp},
\label{eqn28}
\end{equation}
where square brackets indicate projectors, $[a]=\dya{a}$, etc., and
\begin{equation}
  \ket{\Psi_3} = T(t_3,t_0) \ket{\Psi_0} = (\UM\ot\VM) \ket{\Psi_0}.
\label{eqn29}
\end{equation}
One can also write 
\begin{equation}
  \Pr(A,B,a,b) = \mte{\Psi_2}{\Blp [a]\ot P^{(a)}_A\ot Q^{(b)}_B\ot[b]\Brp}
\label{eqn30}
\end{equation}
where $\ket{\Psi_2}=T(t_2,t_0)\ket{\Psi_0}$, and
\begin{equation}
  P^{(a)}_A = {U^{(a)}}\ad [A] U^{(a)},\quad 
 Q^{(b)}_B = {V^{(b)}}\ad [B] V^{(b)}
\label{eqn31}
\end{equation}
are projectors obtain by mapping $[A]$ and $[B]$ backwards in time from $t_3$
to $t_2$, so they correspond to the measurement bases ``chosen'' by the
quantum coin flips.  Since there are no gates between $t_0$ and $t_2$ in
Fig.~\ref{fgr1}(a), one can replace $\ket{\Psi_2}$ in \eqref{eqn30} with
$\ket{\Psi_0}$, and simplify the resulting expression to give
\begin{equation}
  \Pr(A,B,a,b) = \Pr(a)\Pr(b) \mte{\Phi}{P^{(a)}_A\ot Q^{(b)}_B},
\label{eqn32}
\end{equation}
where $\Pr(a)=\mte{\phi_a}{[a]}$ and $\Pr(b)=\mte{\phi_b}{[b]}$.

\xb
\outl{Randomness of measurements}
\xa

Note that for a
\emph{fixed} $a$ $\{P^{(a)}_A\}$ is a decomposition of the $\AM$
identity, 
\begin{equation}
  I_\AM = \sum_A P^{(a)}_A.
\label{eqn33}
\end{equation}
Analogous comments apply to the $Q^{(b)}_B$.  Using \eqref{eqn33} and its
analog for $\BM$ one can sum both sides of \eqref{eqn32} over $A$ and $B$ to
obtain $\Pr(a,b)=\Pr(a)\Pr(b)$, i.e, these variables are statistically
independent, an obvious consequence of our choice of a product initial state
$\ket{\phi_a}\ot\ket{\phi_b}$ in \eqref{eqn27}.  This means that the choice of
measurement on the $\AM$ side is not only random, but also uncorrelated with
the choice of measurement on the $\BM$ side.

\xb
\subsection{Quantum hidden variables}
\label{sct5c}
\xa

\xb
\outl{Assume $\lm$ has to do with properties of particles before measurement}
\xa

Following this somewhat lengthy introduction to the circuit in Fig.~\ref{fgr1}
let us now use it to search for genuine \emph{quantum} properties which might
be counterparts of, or at least resemble in some way, the mysterious $\lm$
that plays a central role in discussions of Bell's inequality in the
literature.  In particular it is interesting to look for some counterpart of
$\lm$ at a time $t_1$ preceding that at which the quantum coins,
Fig.~\ref{fgr1}(b), are flipped to ``choose'' what types of measurements will
be carried out on $\AM$ and $\BM$.  Further let us assume that the quantum
counterpart of $\lm$ is some property or properties of the $\AM$ and $\BM$
particles at this time, as this makes for a relatively simple (though by no
means trivial) discussion, rather than allowing it to also refer to properties
of the ancillary systems or measuring apparatuses.%
\footnote{Bell in his analysis in \cite{Bll90b} introduces another random
  variable $c$, that plays no role in the probabilistic analysis, as it is a
  condition in every probability.  One could use it to denote the apparatus
  and coin states at time $t_1$ in our analysis, without changing any
  conclusion.} %

\xb
\outl{Need histories to describe situation}
\xa

Given that we need a fully quantum mechanical description of the situation and
that the properties of interest are associated with \emph{three} successive
times: $t_0<t_1<t_2$, a quantum histories approach is necessary, as this is
the only known way to carry out a consistent probabilistic discussion in such
a situation within the framework of standard quantum mechanics.  There is, of
course, no sense in which $\lm$ can be ``measured'' in order to produce
probabilities in the usual textbook approach, since we are assuming that the
measurement apparatuses do not interact with the systems of interest until
later.  In the notation introduced in Ch.~8 of \cite{Grff02c} the simplest
quantum sample space of histories associated with Fig.~\ref{fgr1}(a) will be
of the type
\begin{equation}
  [\Psi_0] \od \{\Lm_\lm\} \od \{ [a]\ot P^{(a)}_A\ot Q^{(b)}_B  \ot[b]\}.
\label{eqn34}
\end{equation}
Thus an initial state $[\Psi_0]$ at $t_0$ is followed at $t_1$ by one of the
properties represented by a projector $\Lm_\lm$, where the collection
$\{\Lm_\lm\}$ for different $\lm$, hereafter denoted by $\Lm$, constitutes a
decomposition of the identity $I_{\AM\BM}$, and at $t_2$ we allow the
properties corresponding to $A,B,a,b$ as in \eqref{eqn30}.  (One could also
use the circuit in Fig.~\ref{fgr1}(b); the essential results are the same but
the analysis becomes more complicated because the $[a]$ and $[b]$ events must
be replaced with measurement outcomes.)

\xb
\outl{Interference or decoherence functional }
\xa

The rule for assigning probabilities to such a family of histories can be
written in terms of an interference or decoherence functional \cite{GMHr93}
\begin{equation}
  \JM(A,B,a,b,\lm,\lm') := 
 \mte{\Psi_0}{\Lm_\lm \Blp [a]\ot P^{(a)}_A\ot Q^{(b)}_B  \ot[b]  
 \Brp \Lm_{\lm'}} =
 \dl_{\lm,\lm'} \Pr(A,B,a,b,\lm),
\label{eqn35}
\end{equation}
which is to be interpreted as follows. The first equality defines what is, in
effect, a collection of $m\tm m$ matrices with indices $\lm$ and $\lm'$ in the
case in which the decomposition $\{\Lm_\lm\}$ contains $m$ elements.  For each
possible value of $A$, $B$, $a$, and $b$ there is such a matrix. The second
equality is the \emph{requirement} that each of these matrices be diagonal:
all off diagonal elements, $\lm\neq \lm'$, must vanish, in order for the
interference functional to define a probability.  Provided this
\emph{consistency} (or \emph{decoherence}) condition is satisfied, each
diagonal element defines a probability in the master distribution
$\Pr(A,B,a,b,\lm)$, from which all the others, in particular those that occur
in \eqref{eqn17} and \eqref{eqn18}, can be calculated.

\xb
\outl{There is not a unique choice for $\Lm=\{\Lm_\lm\}$}
\xa

It is important to notice that, just as in classical statistical physics,
there is not a unique choice of sample space, and thus no unique choice for
the projectors in $\Lm=\{\Lm_\lm\}$. The question of interest is therefore
whether there exists some $\Lm$ such that the consistency conditions are
satisfied (as otherwise probabilities are not defined), and such that the
factorization and independence conditions, \eqref{eqn17} and \eqref{eqn18},
are also satisfied.  This requires a consideration of various possibilities,
some of which are taken up in Sec.~\ref{sct5d} below.

\xb
\outl{Formulas for consistency, probabilities, factorization}
\xa

Since we are assuming, as discussed above, that the hidden variables refer to
the $\AM\BM$ subsystem, it is convenient to introduce the states
\begin{equation}
  \ket{\Phi_\lm} = \Lm_\lm\ket{\Phi}
\label{eqn36}
\end{equation}
on $\HM_\AM\ot\HM_\BM$ and rewrite \eqref{eqn35}
in the simpler form
\begin{equation}
  \Pr(a)\Pr(b)\mted{\Phi_\lm}{P^{(a)}_AQ^{(b)}_B}{\Phi_{\lm'}} =
 \dl_{\lm,\lm'} \Pr(A,B,a,b,\lm).
\label{eqn37}
\end{equation}
If consistency is satisfied, summing both sides over $A$ and $B$, and using
using \eqref{eqn33} and its counterpart for $\BM$, lead to the result
\begin{equation}
  \Pr(a,b,\lm) = \Pr(a)\Pr(b)\Pr(\lm),
\label{eqn38}
\end{equation}
where $\Pr(\lm)=\inp{\Phi_\lm}$.  This means that the independence condition
\eqref{eqn18} is satisfied.  Thus we need only consider 
the factorization condition \eqref{eqn17}, which can be written as
\begin{equation}
  \avg{P^{(a)}_AQ^{(b)}_B}_\lm =\avg{P^{(a)}_A}_\lm \avg{Q^{(b)}_B}_\lm,
\label{eqn39}
\end{equation}
where for any operator $R$ on $\HM_\AM\ot \HM_\BM$
\begin{equation}
  \avg{R}_\lm := \mte{\Phi_\lm}{R}/\inp{\Phi_\lm}.
\label{eqn40}
\end{equation}
To summarize up to this point.  We are looking for quantum hidden variables, a
decomposition $\{\Lm_\lm\}$ of the identity on $\HM_\AM\ot \HM_\BM$ which can be
part of a consistent family of histories, that is to say
\begin{equation}
  \mted{\Phi_\lm}{P^{(a)}_AQ^{(b)}_B}{\Phi_{\lm'}} = 0,
   \text{ for } \lm\neq\lm'
\label{eqn41}
\end{equation}
and in addition \eqref{eqn39} holds for all values of $A,B,a,b$ and $\lm$.

\xb
\subsection{Examples}
\label{sct5d}
\xa

\xb
\outl{Many possibilities; we'll only consider a few}
\xa

A full exploration and characterization of quantum hidden variables,
understood as decompositions $\{\Lm_\lm\}$ of $I_{\AM\BM}$ for which the
consistency conditions of families such as \eqref{eqn34} are satisfied, and
hence can be part of a consistent quantum description of the joint probability
distribution of these particles as a function of time, is outside the scope of
the present paper.  Sometimes quantum hidden variables exist satisfying the
factorization and independence conditions \eqref{eqn17} and \eqref{eqn18}, and
sometimes they do not exist; it depends on the initial state $\ket{\Phi}$ of
the $\AM\BM$ system and the types of measurements which are carried out, as
determined by the $U^{(a)}$ and $V^{(b)}$ operators.  We shall only consider a
small number of possibilities, but from these one gets a fairly good idea
of what the issues are.

\xb
\outl{1. von Neumann choice: $\Lm = \{[\Phi], I-[\Phi] \}$ }
\xa

1. The choice 
\begin{equation}
  \Lm = \{[\Phi], I-[\Phi] \},
\label{eqn42}
\end{equation}
that is, a projector onto the initial state $\ket{\Phi}$ together with its
negation, is implicit in von Neumann's discussion of measurement
\cite{vNmn55}, see Sec.~18.2 of \cite{Grff02c}, and hence in much subsequent
work. It leads to $\ket{\Phi_1}= \ket{\Phi}$ and $\ket{\Phi_2} = 0$ in
\eqref{eqn36}.  Consequently the consistency conditions \eqref{eqn37} are
always satisfied. But factorization, a single equation \eqref{eqn39} with
$\lm=1$, will typically not be satisfied.  In such cases we can conclude that
\eqref{eqn42} is \emph{not} a quantum random variable satisfying the
factorization condition.  We cannot thereby conclude that there are no quantum
hidden variables satisfying factorization and independence, but only that if
there are, they are not of the form \eqref{eqn42}.  The reason for this
cautionary note appears in the next example.

\xb
\outl{2. All the $\{P^{(a)}_A\}$ \emph{commute}}
\xa

2. Suppose that for all $a$ and all $A$ the projectors in the collection
$\{P^{(a)}_A\}$ \emph{commute} with each other.  They can then be
simultaneously diagonalized using a single orthonormal basis
$\{\ket{\al_j}\}$, and one can choose as hidden variables the projectors
\begin{equation}
  \Lm_j = [\al_j] \ot I_\BM
\label{eqn43}
\end{equation}
onto these basis states.  The consistency conditions are satisfied and the
states $\ket{\Phi_\lm}$ of \eqref{eqn36} are product states,
\begin{equation}
  \ket{\Phi_j} = \ket{\al_j}\ot\ket{\chi_j},
\label{eqn44}
\end{equation}
so the factorization condition is satisfied: \eqref{eqn39} holds
for every $\lm=j$.  Thus under these circumstances appropriate
quantum hidden variables exist, and Bell's inequality will be satisfied.  Note
that had we used \eqref{eqn42} instead of \eqref{eqn43} the factorization
condition would (in general) \emph{not} hold.  Thus when one choice
of quantum hidden variable fails to satisfy the conditions of factorization
and independence this does not, by itself, exclude the possibility that some
other choice might work better.

The ``mirror image'' of this example, when all projectors in the collection
$\{Q^{(b)}_B\}$ commute with each other, is worth mentioning in that one will
need a different set of quantum hidden variables, $\{I_\AM\ot[\bt_j]\}$
for an appropriate orthonormal basis $\{\ket{\bt_j}\}$ of $\HM_\BM$, to
establish a factorization condition. It is typical of quantum theory that
sample spaces have to be chosen differently in different circumstances, in
contrast to classical physics where one can always in principle employ a
common refinement of the sample spaces of interest, and thus reduce everything
to a single sample space. 

\xb
\outl{3. Singlet state for certain $a$, $b$ violates CHSH; no $\lm$ in this
  case} 
\xa

3. There are, as is well known, instances in which Bell's inequality is
violated.  One of the simplest is when $a$, $b$, $\AM$, $\BM$ are all
two-level systems or qubits, $\ket{\Phi}$ is the singlet state \eqref{eqn2},
the $P^{(a)}_A$ for $a=1$ and 2 project onto the $S_x$ and $S_z$ bases of
$\AM$, and similarly $Q^{(b)}_B$ for $b=1$ and 2 project onto the $S_x$ and
$S_z$ bases of $\BM$. The violation of Bell's inequality in this case proves
there exists no (consistent) choice of hidden variables $\Lm$ for the family
\eqref{eqn34} which will satisfy the factorization condition.

\xb
\outl{Key point: In 2. the projectors commute, and in 3. they don't}
\xa

The contrast between examples 2 and 3 is particularly instructive in analyzing
why a derivation of the Bell-CHSH inequality is sometimes possible and
sometimes fails in the quantum context.  The key to example 2 is that all the
projectors in the collection $\{P^{(a)}_A\}$ \emph{commute} with each other.
It is frequently the case that when operators for physical quantities or
properties that are of interest commute, or at least \emph{almost} commute
with each other, one can successfully get away with classical reasoning
applied to a quantum problem, as in the case of the golf balls in
Sec.~\ref{sct4}. On the other hand, classical reasoning tends to break down in
situations in which noncommutivity is too large to be ignored.  The
commutativity of interest in the present circumstance has to do with
\emph{local} operators referring to properties of $\AM$ (or $\BM$ in the
mirror image of example 2), not nonlocal operators such as the $[\Phi]$ that
projects on an entangled state.  Therefore these examples lend no support to
the idea that quantum mechanics implies mysterious nonlocal effects or
influences.  Instead, it is quantum \emph{incompatibility}, whether local or
nonlocal, that is the central issue.

\xb
\outl{Summary}
\xa

In summary, the factorization and independence conditions \eqref{eqn17} and
\eqref{eqn18} whose conjunction leads to the Bell-CHSH inequality can be
studied from a quantum perspective provided one can find appropriate quantum
properties associated with the hidden variable $\lm$, properties which must
satisfy the consistency conditions in order to have well defined
probabilities. Sometimes hidden variables exist for which \eqref{eqn17} and
\eqref{eqn18} are satisfied, and sometimes they do not exist.  But their
nonexistence is no indication of the existence of long range influences.
Instead, it is one more example of the fact that the quantum world does not
obey the laws of classical physics.

\xb
\section{Genuine Locality}
\label{sct6}
\xa

\xb
\outl{Definition of Einstein locality}
\xa

Let us define \emph{Einstein Locality} in the following way%
\footnote{The wording is essentially the same as in Sec.~2 of \cite{Mrmn98},
  where Mermin refers to it as ``generalized Einstein locality.''  For
  Einstein's own statement see p.~85 of \cite{Enst51}: ``But on one
  supposition we should, in my opinion, absolutely hold fast: the real factual
  situation of the system $S_2$ is independent of what is done with the system
  $S_1$, which is spatially separated from the former.''} %

\begin{quote}

Objective properties of isolated individual systems do not change when 
something is done to another non-interacting system
\end{quote}

\xb
\outl{Explication in terms of slips of paper}
\xa

Before presenting the argument that this is a valid quantum principle let us
again  consider the example of red and green slips of paper
in Sec.~\ref{sct3}.  Alice's opening her envelope, looking at the slip of
paper to see what color it is, and then dropping it into her files or into the
fireplace has absolutely no effect on the properties of the slip of paper in
Bob's envelope, including its color.  It is her \emph{knowledge} of what is in
Bob's envelope, not the \emph{objective} properties of that slip of paper, that
changes.  Spacelike separation in the relativistic sense is not required; it
is only necessary that the one envelope or its contents not interact with the
other.  One could, for example, imagine a very sophisticated arrangement in
which one envelope is equipped with a radio transmitter and the other a
receiver coupled to a device which releases a dye that changes the color of
the slip of paper.  Placing the two envelopes at a sufficient spacelike
separation from each other during the time interval of interest would, of
course, prevent that sort of signaling. but relativity is not the real issue.

\xb
\outl{Quantum objective properties require framework where they are defined}
\xa

Quantum theory describes properties of a system using (projectors on)
subspaces of the appropriate Hilbert space, and as noted in Sec.~\ref{sct2},
incompatible subspaces cannot be combined to yield meaningful
properties. Hence a meaningful quantum discussion of objective properties of
an isolated system requires the use of a framework containing those
properties (or their projectors), and this will necessarily exclude other
incompatible frameworks from the discussion. The very concept of Einstein
locality requires frameworks in which those properties of the isolated system
we are concerned about have some meaning, as otherwise it is obviously
impossible to discuss whether they do or do not change.

\xb
\outl{Isolated Qm systems $\lra$ factorization of dynamics}
\xa

A quantum system $\AM$ can be said to be \emph{isolated} from another system
$\BM$ provided the overall time development operator for the two systems is of
the form $T_\AM(t',t)\ot T_\BM(t',t)$ during the time interval of interest.
Deciding whether an objective property of $\AM$ will change when something
is done to $\BM$ is best discussed using a third system $\CM$ that ``does''
the something, which is to say it interacts with $\BM$ but not with $\AM$.
Thus we assume an overall dynamics of the tripartite system of the form
\begin{equation}
  T(t',t) = T_\AM(t',t)\ot T_{\BM\CM}(t',t).
\label{eqn45}
\end{equation}
And in order to study the effects of what $\CM$ does to $\BM$ we
assume an initial state of the form
\begin{equation}
  \ket{\Psi_0}= \ket{\Phi}_{\AM\BM} \ot \ket{\phi}_\CM
\label{eqn46}
\end{equation}
at $t_0$, and ask whether properties of $\AM$ are altered by using
different initial states $\ket{\phi}$ for $\CM$.  

\xb
\outl{Joint probabilities for histories of the $\AM$ system}
\xa

Consider a particular sequence of properties $A_j^{(\al_j)}$ of $\AM$ at a
succession of times $t_j$ during which it is isolated, so \eqref{eqn45}
applies.  The joint probability distribution can be calculated using the
interference or decoherence functional \cite{GMHr93}---see also see Sec.~10.2
of \cite{Grff02c} and compare with \eqref{eqn35}---
\begin{equation}
  \JM(\al_1,\al_2\ldots,\al'_1,\al'_2\ldots) =
  \mte{\Psi_0}{K(\al)\ad K(\al')} = \Tr_\AM[\rho_\AM K(\al)\ad K(\al')]
\label{eqn47}
\end{equation}
where $\al$ stands for $\al_1,\al_2,\ldots$, the chain operator $K(\al)$ is
defined by
\begin{equation}
  K(\al) := A_n^{(\al_n)}T(t_{n},t_{n-1})A_{n-1}^{(\al_{n-1})} \cdots 
  A_1^{(\al_1)} T(t_1,t_0), \quad
\label{eqn48}
\end{equation}
and 
\begin{equation}
\rho_\AM := \Tr_{\BM\CM}(\dya{\Psi_0})
 = \Tr_\BM( \dya{\Phi})  
\label{eqn49}
\end{equation}
is the reduced density operator of $\AM$ at $t_0$.  If the interference
functional $\JM(\al,\al')$ vanishes whenever $\al_j\neq \al'_j$ for some $j$,
which is to say the consistency conditions are satisfied, its diagonal
elements provide the probabilities of the corresponding histories of the $\AM$
system.  The point is that \emph{neither consistency nor the resulting
  probabilities depend in any way on the choice of the initial state
  $\ket{\phi_\CM}$ for system $\CM$}, because $\rho_\AM$, \eqref{eqn49}, is
independent of $\ket{\phi_\CM}$.  This completes the argument that whatever is
done to the system $\BM$ has absolutely no effect upon any of the properties,
or a time sequence of such properties, of the non-interacting system $\AM$.
The reader who prefers to see how things work out in a particular instance,
rather than relying on the general proof given here, will find an example in
Sec.~23.4 in \cite{Grff02c}.

\xb
\section{Conclusion}
\label{sct7}
\xa

\xb
\outl{Absence of nonlocal influences in QM shown in two ways}
\xa

The basic conclusions of this paper can be summarized as follows.  Although
quantum theory involves the use of nonlocal \emph{states}, such as wave
packets and entangled states, there is nothing in the theory, or in the real
world so far as it is accurately described by quantum theory, that corresponds
to the sorts of instantaneous nonlocal \emph{influences} which have often been
thought to arise in the situation envisaged in the EPR paradox, or implied by
the fact that quantum theory violates Bell inequalities.  When quantum
mechanics is consistently applied using properly formulated probabilities on
properly defined sample spaces the complete absence of such nonlocal
influences can be shown in two ways.

\xb
\outl{1. Defects in derivations of Bell inequalities}
\xa

First, derivations of Bell inequalities, while valid for all practical purposes
in a macroscopic situation in which the appropriate quantum projectors
commute, fail when the classical approximations to quantum theory are no
longer valid.  Therefore it is not surprising that they are violated by
quantum theory as applied to suitable microscopic objects.  What this
violation tells us is not the locality breaks down, but rather that classical
physics no longer applies in the quantum domain.  To be sure, the analysis in
this paper only applies to some particular derivations of these inequalities,
but there is no reason to believe that others escape the fundamental
difficulties which we have identified: the unjustified use of classical ideas,
in particular the careless use of probabilities, in the quantum domain.

\xb
\outl{2. Einstein locality is good QM; not surprising given that supposed
  superluminal influences cannot carry information}
\xa

Second, the same techniques that reveal deficiencies in derivations of Bell
inequalities can be used to demonstrate in a positive sense a principle of
Einstein \emph{locality} which directly contradicts any notion of nonlocal
influences.  That such a principle is correct is hardly surprising given that
even the most enthusiastic proponents of nonlocal influences have conceded
that these cannot be used to transmit signals or information.  The advance
represented by the formulation used in the present paper is that confusing
ideas and possible loopholes associated with an ill-defined concept of
``measurement''  are absent from the discussion, and no longer provide a
screen behind which such supposed influences can conceal their nonexistence.
Nonlocal influences cannot be used to transmit signals for the simple reason
that there are none.  

\xb
\outl{Spurious nonlocal influences arise from mistaken use of Cl reasoning}
\xa

Our results indicate that spurious nonlocal influences arise from mistaken
reasoning, from applying classical modes of analysis in the quantum domain
where they do not apply.  While the fundamental source of the problem was
pointed out some time ago by Fine \cite{Fne82}, the tools for dealing with it
decisively using a fully consistent formulation of quantum probabilities have
only been developed more recently. These allow a much more precise discussion
of the problem than was possible in the past. 

\xb
\outl{Reconciling QM and SR has nothing to do with nonlocal influences}
\xa

Whatever difficulties may remain in reconciling quantum mechanics and special
relativity have nothing to do with the spurious nonlocal influences thought to
arise because of the mistaken application of Bell inequalities to the quantum
domain. Certain \emph{alternatives} to standard Hilbert space quantum theory,
in particular the de Broglie Bohm pilot wave approach \cite{Glds06} and the
spontaneous localization scheme of Ghirardi, Rimini and Weber \cite{BsGh03},
are known to possess a basic incompatibility with special relativity.
However, standard quantum mechanics when give a consistent stochastic
interpretation does \emph{not} suffer from these problems, contrary to claims
in some popular (e.g., \cite{AlGl09}) as well as more technical (e.g.,
\cite{Mdln02}) discussions. (For more details on how probabilities can be
consistently introduced into relativistic quantum mechanics see
\cite{Grff02b}.)

\xb
\outl{What is wrong with ``local realism''}
\xa

The analysis in this paper implies that claims that quantum theory violates
``local realism'' are misleading. To be sure, quantum mechanics allows
entangled states which are incompatible with, as discussed in Sec.~\ref{sct2},
certain local properties.  That this disagrees with intuitions based upon
pre-quantum physics is obvious, but why refer to the latter as ``realistic''?
It is the quantum world, not the classical world, which is the real world
according to modern physics. And quantum mechanics is in full accord with the
principle of Einstein locality stated in Sec.~\ref{sct6}.  Now it is true that
a science can survive despite misleading terminology---``heat capacity'' in
modern thermodynamics, in which heat is no longer regarded as a fluid, comes
to mind as one example.  Nonetheless, there is much to be gained by adopting
terms which reflect rather than stand in disagreement with our best current
scientific understanding of a subject.  It would be quite appropriate to say
that quantum theory contradicts ``\emph{classical} local realism,'' or, better
yet, ``\emph{classical} realism.''  It is classical realism, not local
realism, that is in serious disagreement with the best experimental results
currently available.

\xb
\outl{Qm incompatibility more interesting than spurious nonlocality}
\xa

One wonders whether the energy expended discussing nonlocality might not be
better used for investigating what is genuinely mysterious and surprising
about quantum theory in contrast to classical physics: \emph{quantum
  incompatibility}.  Whereas entangled states that are incompatible with local
properties provide one manifestation of this, there are many others. The
subject is not well understood, either at the mathematical or the intuitive
level, and is nowadays an area of active research in the field of quantum
information.  Thus it is incompatibility rather than nonlocal influences that
deserves attention by those interested in the foundations of quantum
mechanics.  We have a situation in which one can honestly say that truth is
stranger, but also a lot more interesting, than fiction.

\section*{Acknowledgments}
I am grateful for extensive comments by an anonymous referee. The research
described here received support from the National Science Foundation through
Grant 0757251.

\xb


\end{document}